\documentclass[twoside,11pt]{article}

\usepackage{blindtext}

\usepackage[abbrvbib, preprint]{jmlr2e}
\usepackage{booktabs}
\usepackage{multirow}
\usepackage{amsmath}

\usepackage{listings}
\usepackage{xcolor}

\definecolor{codegreen}{rgb}{0,0.6,0}
\definecolor{codegray}{rgb}{0.5,0.5,0.5}
\definecolor{codepurple}{rgb}{0.58,0,0.82}
\definecolor{backcolour}{rgb}{0.95,0.95,0.92}

\lstdefinestyle{mystyle}{
    backgroundcolor=\color{backcolour},   
    commentstyle=\color{codegreen},
    keywordstyle=\color{magenta},
    numberstyle=\tiny\color{codegray},
    stringstyle=\color{codepurple},
    basicstyle=\ttfamily\footnotesize,
    breakatwhitespace=false,         
    breaklines=true,                 
    captionpos=b,                    
    keepspaces=true,                  
    showspaces=false,                
    showstringspaces=false,
    showtabs=false,                  
    tabsize=2
}

\usepackage{lastpage}
\jmlrheading{23}{2026}{1-\pageref{LastPage}}{1/21; Revised 5/22}{9/22}{21-0000}{Colombe Becquart}

\ShortHeadings{ICSpyLab: A Python Package for ICS}{Becquart}
\firstpageno{1}

\begin{document}

\title{ICSpyLab: A Python Package for Invariant Coordinate Selection }

\author{\name Colombe Becquart \email colombe.becquart@tse-fr.eu \\
       \addr Toulouse School of Economics, \\ Université Toulouse Capitole \\
       France\\
       Université de Toulouse\\
       France
       }

\editor{My editor}

\maketitle

\begin{abstract}
Invariant coordinate selection (ICS) is a dimensionality reduction technique based on the joint diagonalization of two scatter matrices. While principal component analysis relies solely on variance, ICS seeks directions of maximal or minimal generalized kurtosis, making it a powerful alternative for clustering and anomaly detection. Despite its theoretical and practical relevance, no dedicated Python implementation of ICS is currently available, limiting its integration into modern machine-learning workflows.
We introduce \texttt{ICSpyLab}, the first Python package implementing ICS. It provides a broad collection of scatter matrices, multiple algorithms for computing invariant components, and several component-selection criteria. Designed for both practitioners and researchers, \texttt{ICSpyLab} follows a standard estimator interface, facilitating its use within machine-learning pipelines while remaining flexible for methodological extensions. The documentation includes detailed explanations and reproducible examples. \texttt{ICSpyLab} is released under the MIT license and is openly available at: \url{https://github.com/cbecquart/ICSpyLab}.
 
\end{abstract}


\section{Introduction}

Dimensionality reduction plays a central role in modern statistical learning, providing low-dimensional representations that facilitate interpretation, visualization, and downstream tasks such as clustering and anomaly detection \citep{jia_feature_2022}. While principal component analysis (PCA) remains the dominant approach, its reliance on second-order moments limits its ability to capture deviations from Gaussianity or to reveal structure driven by higher-order moments.

Invariant coordinate selection \citep[ICS;][]{tyler_invariant_2009, nordhausen2022usage, archimbaud2026review} is an alternative dimensionality reduction technique based on the joint diagonalization of two scatter matrices. A scatter matrix is a positive definite matrix that summarizes the dispersion of a multivariate random vector and generalizes the notion of a covariance matrix. It is uniquely determined by the underlying distribution and satisfies an affine equivariance property. By exploiting differences in scatter estimates, ICS identifies directions associated with extreme values of generalized kurtosis. ICS exhibits promising theoretical properties, including affine invariance and links to Fisher’s discriminant subspace under some mixture models \citep{tyler_invariant_2009, becquart2025invariant}. As a comparison, PCA seeks directions of maximal variance by diagonalizing one scatter matrix, and the resulting components are only orthogonal invariant \citep{jolliffe_principal_2002}.
Recent work has demonstrated the outperformance of ICS as a preprocessing step in several clustering \citep{fischer2017subgroup, alfons_tandem_2024} and anomaly detection \citep{archimbaud_unsupervized_2018} applications.

Despite its theoretical appeal and successful applications, established software implementations of ICS are limited to the R ecosystem, split through the packages \texttt{ICS} \citep{nordhausen_tools_2008}, \texttt{ICSOutlier} \citep{archimbaud_unsupervized_2018}, and \texttt{ICSClust} \citep{archimbaud_icsclust_2023}. While these packages contain a variety of ICS functionalities, they are not accessible to Python users. In addition, these implementations are not tailored to modern machine learning workflows and lack cohesion, as the full ICS pipeline is spread across three independent packages.
To address this gap, we introduce \texttt{ICSpyLab}, the first Python library to provide an implementation of ICS, drawing on the existing R packages. The package consolidates the key components of these R libraries within a unified framework adapted to the Python environment. 
The package provides a user-friendly, scikit-learn-compatible API \citep{buitinck_api_2013}, allowing streamlined execution of the ICS workflow. It includes multiple scatter matrix options, four algorithms for computing ICS, and several component-selection criteria. \texttt{ICSpyLab} also supports user-defined scatter matrices and component-selection methods. Such use cases, along with additional examples, are provided in the extensive documentation, thereby facilitating experimentation. The project is open-source and released under the MIT license. Contributions of every kind are welcome and contribution guidelines are available in the GitHub repository: \url{https://github.com/cbecquart/ICSpyLab}.

\section{Package Implementation}

\paragraph{Features.} In order to apply ICS, it is necessary to select the following parameters: two scatter matrices for which the joint diagonalisation is desired, an algorithm for calculating invariant components, and a component selection method (which can be omitted to retain all components). Table~\ref{tab:features} summarizes the features currently implemented in \texttt{ICSpyLab} for each of these parameter categories. The package also provides additional utilities for data generation and visualization, including generators for multivariate exponential power distributions and mixtures of elliptical distributions.

\begin{table}[ht!]
\centering
\caption{Main functionalities currently implemented in \texttt{ICSpyLab}.}
\begin{tabular}{lll}
\toprule
\textbf{Category} & \textbf{Feature} & \textbf{Reference} \\
\midrule

\multirow{4}{*}{ICS algorithm}
& \texttt{eigh} & \cite{virtanen_scipy_2020} \\ 
& \texttt{standard} & \cite{nordhausen_tools_2008} \\
& \texttt{whiten}   & \cite{nordhausen_tools_2008} \\
& \texttt{QR}       & \cite{archimbaud2023numerical} \\

\midrule

\multirow{8}{*}{Scatter matrix}
& \texttt{cov}       &  Classical covariance \\
& \texttt{covW}      & \cite{tukey1977exploratory} \\
& \texttt{cov4}      &  \cite{Cardoso1989} \\
& \texttt{covAxis}   &  \cite{critchley_principal_2008} \\
& \texttt{mcd}       & \cite{rousseeuw_least_1984} \\
& \texttt{tcov}      & \cite{caussinus1995metrics} \\
& \texttt{tcovAxis}  & \cite{tyler_invariant_2009} \\
& \texttt{tM}        & \cite{kent_curious_1994} \\

\midrule

\multirow{3}{*}{Component selection}
& Median criterion      & \cite{alfons_tandem_2024} \\
& Normal criterion   & \cite{alfons_tandem_2024} \\
& Unimodality criterion & \cite{becquart2026note} \\

\bottomrule
\end{tabular}
\label{tab:features}
\end{table}

\paragraph{API design.} The core component of \texttt{ICSpyLab} is the \texttt{ICS} class, which inherits from scikit-learn’s \texttt{BaseEstimator}. This design choice ensures compatibility with the scikit-learn ecosystem, including pipelines and cross-validation tools. In contrast to existing R implementations, the Python API follows the conventional estimator interface. The \texttt{fit} method estimates the transformation matrix defining the invariant coordinate system, while the \texttt{transform} method applies this transformation to new data once the model has been fitted. For convenience, a \texttt{fit\_transform} method is also provided to perform both steps in a single call.

\paragraph{Extensibility.} Another key component of the package is the \texttt{Scatter} class. Since ICS relies on the joint diagonalization of two scatter matrices chosen from a broad family, this abstraction was introduced to make the framework extensible and modular. It allows users to easily define and integrate new scatter estimators into the ICS pipeline even though several scatter estimators are implemented directly in the package. The same logic applies to the component selection step, through the \texttt{ComponentSelect} class. 

\paragraph{Computational considerations.} The implementation is built on standard Python libraries such as
\texttt{NumPy} \citep{harris_array_2020}, \texttt{Numba} \citep{lam_numba_2015}, \texttt{SciPy} \citep{virtanen_scipy_2020}, \texttt{Matplotlib} \citep{hunter_matplotlib_2007} and
\texttt{scikit-learn} \citep{pedregosa_scikit-learn_2011}, ensuring reliability and consistency with established tools.
The \texttt{mcd} scatter is provided as a wrapper around the scikit-learn implementation. Pairwise scatter matrices (\texttt{tcov} and \texttt{tcovAxis}) are computationally optimized using JIT compilation via \texttt{Numba}, improving computational efficiency and facilitating their use on larger datasets.

\section{Example of package usage}

The following example illustrates the basic usage of \texttt{ICSpyLab} on the RANDU dataset \citep{marsaglia1968random}, a well-known example where data points are trapped in parallel hyperplanes.
Invariant components are computed using the \texttt{fit\_transform} method.

\begin{lstlisting}[language=Python]

from icspylab import ICS
from icspylab.distributions import generate_randu

X = generate_randu()
ics = ICS(S1="cov", S2="tcovAxis", algorithm="standard")
X_ics = ics.fit_transform(X)
\end{lstlisting}

\begin{figure}[htbp]
    \centering
    \begin{minipage}{0.48\textwidth}
        \centering
        \includegraphics[width=\textwidth]{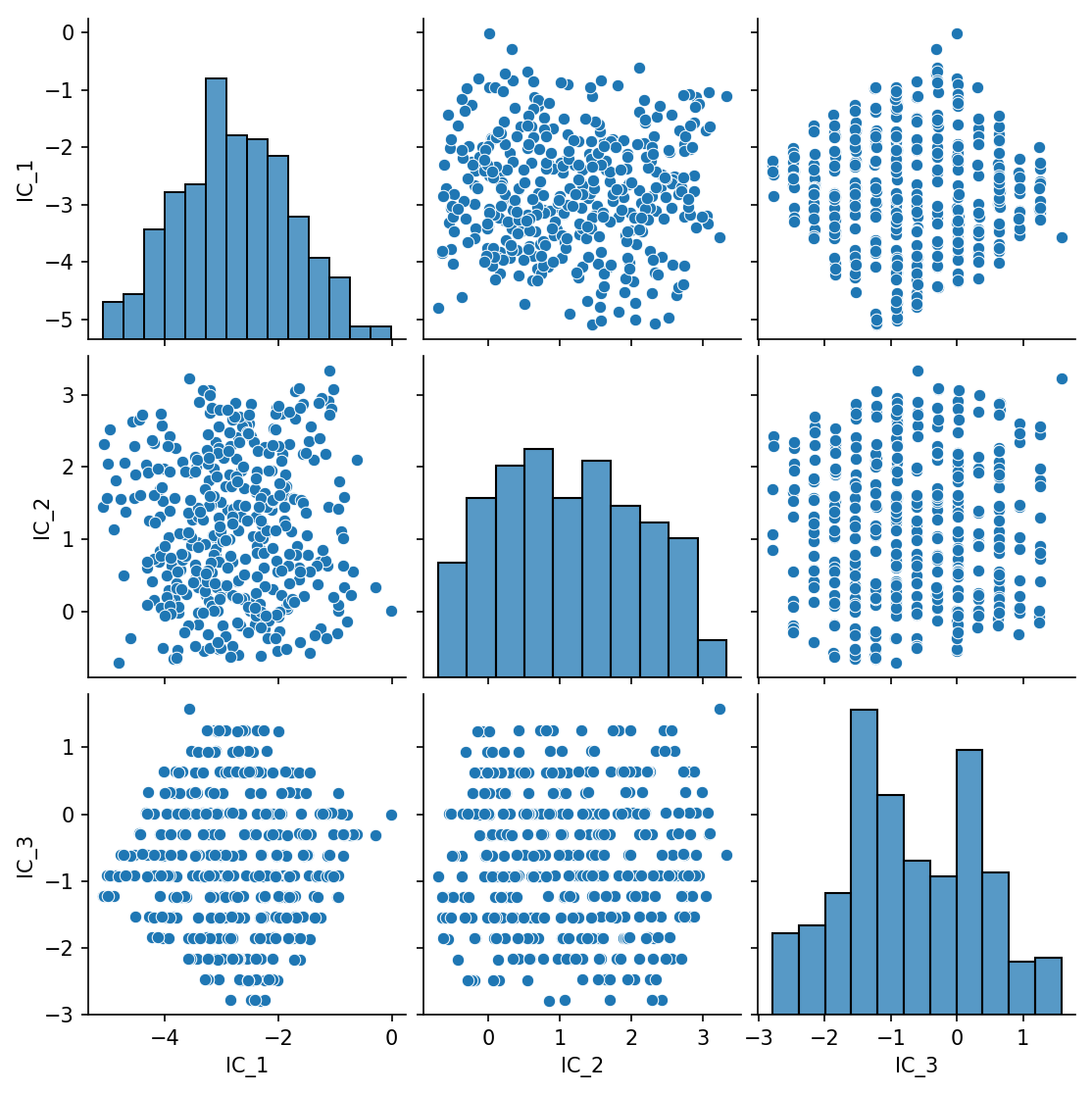}
        \text{(a) Invariant components (ICS).}
        \label{fig:randu_ics}
    \end{minipage}
    \hfill
    \begin{minipage}{0.48\textwidth}
        \centering
        \includegraphics[width=\textwidth]{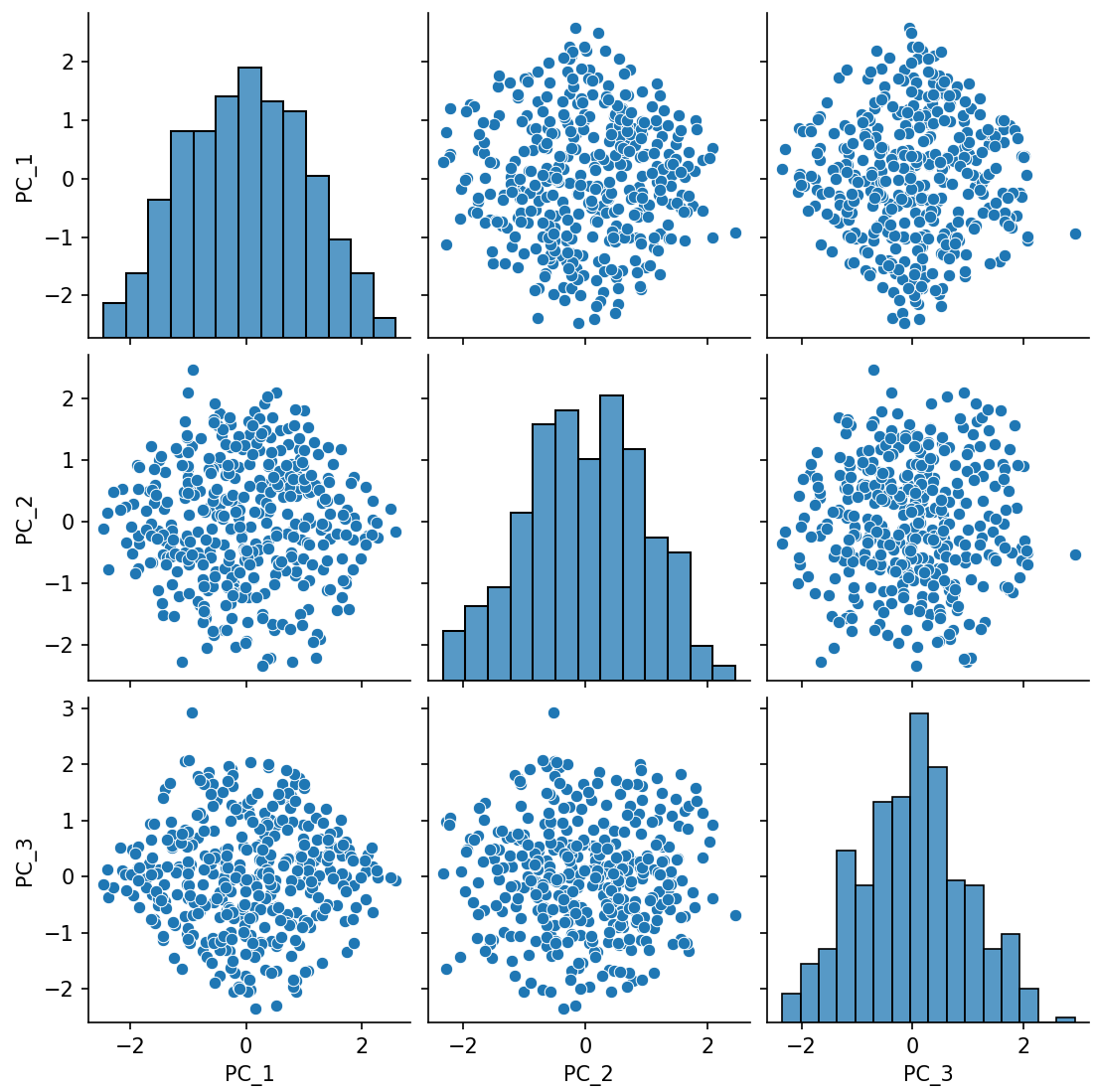}
        \text{(b) Principal components (PCA).}
        \label{fig:randu_pca}
    \end{minipage}
    \caption{Projections of the RANDU dataset: (a) invariant components computed via \texttt{ICSpyLab} revealing the hyperplane alignments, and (b) principal components.}
    \label{fig:randu_comparison}
\end{figure}

Figure~\ref{fig:randu_comparison} illustrates how ICS reveals the underlying hyperplane structure on the last invariant component, whereas principal components computed by PCA fail to do so. The complete experiment and plotting scripts are detailed in the package documentation.

\section{Conclusion and Future Work}

\texttt{ICSpyLab} brings ICS to the Python environment through a unified and extensible implementation, making the method more accessible for practical applications and methodological research.

Future development will focus on strengthening the collaborative and open-source nature of the project, with the goal of fostering contributions from an active and rapidly growing Python-based scientific community. 
Ongoing research directions within the ICS community open promising avenues for expansion, including adaptations to functional data, the development of new component selection criteria, and extensions tailored to high-dimensional settings.

\acks{  }
The author would like to thank Abdallah Abdelsameia for his contributions to the implementation of \texttt{ICSpyLab} during his internship, and the attendants of the 2026 ``ICS and related methods" workshop for stimulating discussions.

\vskip 0.2in
\bibliography{references}

\end{document}